%% file: main.tex
\documentclass[sigconf]{acmart}
\usepackage{xcolor} 

\AtBeginDocument{%
  \providecommand\BibTeX{{%
    \normalfont B\kern-0.5em{\scshape i\kern-0.25em b}\kern-0.8em\TeX}}}

\setcopyright{CC}
\setcctype[4.0]{by-nc}

\copyrightyear{2025}
\acmYear{2025}
\acmConference[ISWC '25]{Proceedings of the 2025 ACM International Symposium on Wearable Computers}{October 12--16, 2025}{Espoo, Finland}
\acmBooktitle{Proceedings of the 2025 ACM International Symposium on Wearable Computers (ISWC '25), October 12--16, 2025, Espoo, Finland}\acmDOI{10.1145/3715071.3750405}
\acmISBN{979-8-4007-1481-8/2025/10}

\usepackage{xac}
\usepackage{booktabs}
\usepackage{multirow}
\usepackage{array}
\usepackage{makecell}
\usepackage{caption}
\usepackage{siunitx}
\usepackage{placeins}

\usepackage{textgreek}
\usepackage{scalerel}

\makeatletter
\def\@seccntformat#1{\@ifundefined{#1@cntformat}%
   {\csname the#1\endcsname\quad}  
   {\csname #1@cntformat\endcsname}
}
\let\oldappendix\appendix 
\renewcommand\appendix{%
    \oldappendix
    \newcommand{\section@cntformat}{\MakeUppercase{\appendixname~\thesection}\quad}
}
\makeatother

\begin{document}

\title{EchoForce: Continuous Grip Force Estimation from Skin Deformation Using Active Acoustic Sensing on a Wristband}

\author{Kian Mahmoodi}
\authornote{Authors contributed equally.}
\orcid{0009-0009-7660-1922}
\email{km777@cornell.edu}
\affiliation{%
  \institution{Cornell University}
  \city{Ithaca}
  \state{NY}
  \country{USA}}

\author{Yudong Xie}
\authornotemark[1]
\orcid{0009-0003-1409-7827}
\email{xyd22@mails.tsinghua.edu.cn}
\affiliation{%
  \institution{Tsinghua University}
  \city{Beijing}
  \country{China}}

\author{Tan Gemicioglu}
\authornotemark[1]
\orcid{0000-0001-9324-4431}
\email{tg399@cornell.edu}
\affiliation{%
  \institution{Cornell Tech}
  \city{New York}
  \state{NY}
  \country{USA}}

\author{Chi-Jung Lee}
\orcid{0000-0002-1887-4000}
\email{cl2358@cornell.edu}
\affiliation{%
  \institution{Cornell University}
  \city{Ithaca}
  \state{NY}
  \country{USA}}

\author{Jiwan Kim}
\orcid{0000-0002-0806-2797}
\email{jiwankim@kaist.ac.kr}
\affiliation{
    \institution{KAIST}
    \city{Daejeon}
    \country{Republic of Korea}
}

\author{Cheng Zhang}
\orcid{0000-0002-5079-5927}
\email{chengzhang@cornell.edu}
\affiliation{%
  \institution{Cornell University}
  \city{Ithaca}
  \state{NY}
  \country{USA}}


\begin{abstract}
  \input{00_abstract}
\end{abstract}




\begin{teaserfigure}
  \centering
  \vspace{-1em}
  \includegraphics[width=\textwidth]{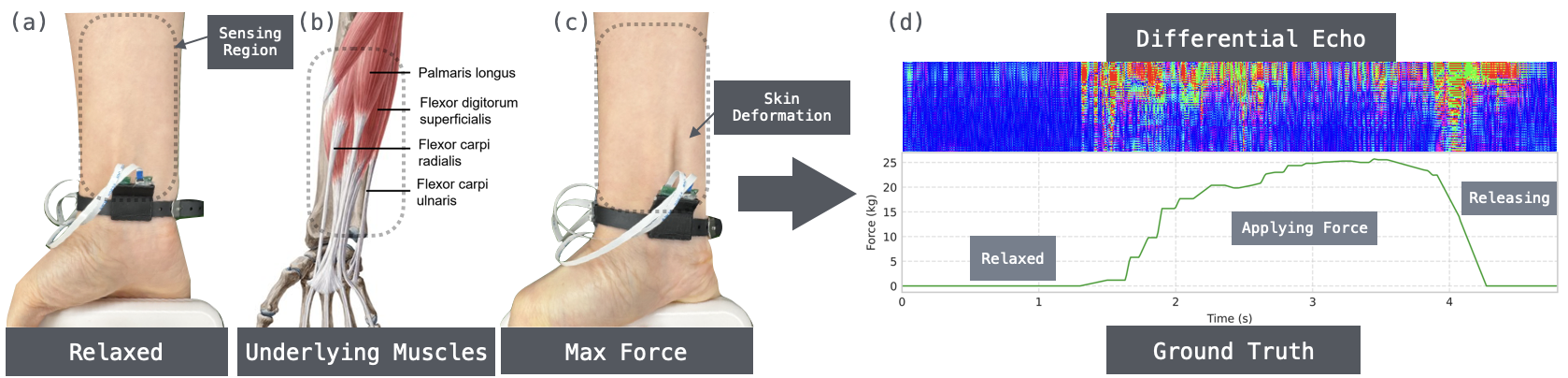}
  \caption{
  Overview of EchoForce’s sensing approach. (a) Wristband on a relaxed forearm. (b) Flexors of the anterior forearm, which cause distinctive skin deformations. (c) Wristband under maximum voluntary grip force, showing pronounced surface deformation. (d) Differential echo profile from the sensor and the corresponding grip force. Sensing area marked in gray.
  }
  \Description{Overview of EchoForce’s sensing approach. (a) Photograph of the wristband on a relaxed forearm. (b) Flexor muscles of the anterior forearm close to the skin surface, which cause distinctive skin deformations. (c) Photograph of the same wristband under maximum voluntary grip force, showing pronounced surface deformation. (d) The differential echo profile from the sensor and the corresponding grip force as measured by a dynamometer. The sensing area is marked in gray.}
  \label{fig:teaser}
\end{teaserfigure}


\maketitle

\input{01_introduction}

\input{02_relatedwork}

\input{03_system}
\input{04_study}

\input{05_results}
\input{06_discussion}

\input{07_conclusion}
\input{08_acknowledgements}

\bibliographystyle{ACM-Reference-Format}
\bibliography{ref_main}

\appendix

\end{document}

%% file: 00_abstract.tex
Grip force is commonly used as an overall health indicator in older adults and is valuable for tracking progress in physical training and rehabilitation. Existing methods for wearable grip force measurement are cumbersome and user-dependent, making them insufficient for practical, continuous grip force measurement. We introduce EchoForce, a novel wristband using acoustic sensing for low-cost, non-contact measurement of grip force. EchoForce captures acoustic signals reflected from subtle skin deformations by flexor muscles on the forearm. In a user study with 11 participants, EchoForce achieved a fine-tuned user-dependent mean error rate of 9.08\% and a user-independent mean error rate of 12.3\% using a foundation model. Our system remained accurate between sessions, hand orientations, and users, overcoming a significant limitation of past force sensing systems. EchoForce makes continuous grip force measurement practical, providing an effective tool for health monitoring and novel interaction techniques.

%% file: 01_introduction.tex
\section{Introduction}

Grip force is a measurement of the contact forces applied by the hand, serving as a standardized method to assess muscle strength. Grip force is commonly used to evaluate overall health among older adults \cite{celis-morales_associations_2018}. Grip force can also be used for monitoring progress in strength training and physical rehabilitation \cite{kurillo_grip_2005}. In HCI, grip forces have been used to provide novel input methods independent of the object or surface the hand is in contact with \cite{saponas_enabling_2009,buddhika_fsense_2019}. 

While grip force measurement is useful for a wide range of clinical and commercial applications, current methods for measuring grip force are cumbersome, and not suitable for continuous, wearable usage. The most common approach for measuring grip force precisely is specialized handheld devices such as grip dynamometers, which fully occupy users' hands during use \cite{sanchez-aranda_reliability_2024}. Wearable approaches have used force-sensitive resistors (FSRs) and electromyography (EMG) as an alternative to measure grip force. FSRs are placed on the fingertip for force measurement and may restrict joint mobility \cite{lin_novel_2020}, whereas EMG typically needs expertise to setup, is expensive, and may become inaccurate if mispositioned over the forearm \cite{merlettiElectromyography2004}. Moreover, both systems are highly sensitive to changes in environment, user, or session, requiring calibration and adjustments whenever any significant change is made. To make continuous grip force measurement practical for health monitoring and human-computer interaction, there is a significant need for an inexpensive, easy-to-use wearable device that can adapt to new sessions or users without calibration.

EchoForce is a compact, low-cost wristband for continuously measuring grip force by tracking skin deformations on the forearm with active acoustic sensing, designed to fill a gap in the generalizability of past grip force sensing approaches. The device emits inaudible acoustic signals towards the forearm, which echo back from the skin. The echoes are then captured by the microphones on the wristband. EchoForce's sensing approach is enabled by the unique positioning of finger flexors. Finger muscles, such as the \textit{flexor digitorum superficialis}, are located in the forearm and connected to the fingers by tendons. When the flexor muscle is tensed and the tendon is pulled, this results in skin deformations around the wrist as shown in Figure \ref{fig:teaser}c. By capturing these subtle skin deformations with active acoustic sensing, we can estimate how much force is being applied by the finger muscles.

EchoForce uses a small speaker and microphone for sensing, making it comfortable to wear and easy to integrate force sensing into wrist-based wearables. It can reliably measure grip force without needing direct contact with skin, unlike previous wearable force sensing techniques that can lead to discomfort. Using a fine-tuned foundation model, EchoForce achieves mean 9.08\% user-dependent error rate -- comparable with EMG -- independent of remounting sessions and wrist orientation. Moreover, the system achieves a mean error rate of 12.3\% in the user-independent model, increasing the generalizability of our sensing approach to new users, which is known to be a key limitation for existing methods (e.g., EMG-based). 


%% file: 02_relatedwork.tex
\section{Related Work}

\subsection{Grip Force Estimation}
Assessing grip strength has been a key component of physical therapy and physiological assessments since the 1950s, starting with Bechtol's hydraulic dynamometer, which has remained the clinical standard \cite{bechtol_grip_1954}. Since then, various sensing methods have been proposed to overcome the need to hold a dynamometer in hand. Early work on surface electromyography (sEMG) showed a relationship between electrical signals produced by muscle activity and force application \cite{maier_emg_1995}, which was used to estimate grip force directly \cite{keir_development_2005,hoozemans_prediction_2005}. Further work on sEMG showed its extensibility to multiple degrees of freedom \cite{nielsen_simultaneous_2011,liu_emg-force_2013, forman_influence_2019}, but also that it may have increased error when fatigued \cite{soo_estimation_2010}. When deployed in real-world settings, sEMG-based methods are faced with recalibration issues between sessions and users \cite{hashemi_surface_2013}, high noise \cite{CLANCYSampling20021, PhinyomarkFeatureExtraction2012}, sensitive sensor placement \cite{merlettiElectromyography2004}, and complex and non-portable devices \cite{chowdhurySurfaceElectromyography2013}.

Recently, estimating grip and contact forces has gained attention in HCI for enabling always-available interaction, especially for augmented reality \cite{saponas_demonstrating_2008,mollynEgotouch2024}. Computer vision has been used to estimate grip forces with 10.1\% error \cite{jeongHandgrip2021} and 6.8\% error for minor contact forces \cite{mollynEgotouch2024}. HIPPO uses light reflectivity to achieve 4.2kg error when integrating user-dependent information \cite{yin_hippo_2023}. However, both light-based systems may struggle with occlusions and varying lighting conditions. Sakuma et al. use accelerometers and nail deformation to measure contact forces, but the device restricts finger joint mobility and is not suitable for measuring large grip forces \cite{sakuma_wearable_2018}. Rudolph et al. used capacitive sensors on the wrist to utilize the skin deformation principle used by EchoForce, but found a high regression error of 27.9\% \cite{rudolph_sensing_2022}. Funato and Takemura used a bone conduction-based active acoustic sensing method where the arm was vibrated and the vibrations were physically propagated through the body \cite{funato_grip_2017}. In contrast, EchoForce uses a speaker to emit ultrasound, and a microphone to capture the altered, reflected signals from skin surface deformations, enabling a non-contact, distraction-free approach suitable for continuous wearable use.

\subsection{Active Acoustic Sensing} 
Active acoustic sensing generates encoded acoustic signals and analyzes their direct transmission or reflected echoes from surrounding environments with microphones. Acoustic sensing can be done using only speakers and microphones that are widely integrated into computing devices without needing any modification, making it easy to integrate into existing systems \cite{sonarID22, lipwatch24,kim_cross_2025}. Acoustic signals can travel through contact surfaces \cite{ono_touch_2013} or the body \cite{funato_grip_2017,zhang2017bioacoustics,zhang2018fingerping}. EchoForce instead uses its reflection, which has recently shown promise for bodily sensing through broad applications including activity recognition \cite{mahmudActSonic2024,zhangEatingTrak2022, parikhEchoGuide2024, mahmud2024munchsonic}, pose tracking \cite{mahmudPoseSonic2023,lee2024echowrist}, and physiological sensing \cite{sunEchoNose2023, guo_echobreath_2025}. EchoForce applies active acoustic sensing to grip force estimation, providing a low-cost, compact, and user-independent device for accurately tracking grip forces.

%% file: 03_system.tex
\section{System Design}

\subsection{Sensing Principle} \label{sensing_principle}

Grip forces are exerted by flexing muscles in the anterior forearm, including the flexor digitorum superficialis, flexor digitorum profundus, flexor pollicis longus, and palmaris longus. When these muscles are activated, they pull on tendons that insert into the fingers. Tendons attached to the muscles closer to the skin, the flexor digitorum superficialis and palmaris longus, lead to pronounced skin deformations on the wrist, as shown in Figure \ref{fig:teaser}c.

To capture these subtle yet distinct deformations, we adopted an active acoustic sensing system for fine-grained around-movement measurement~\cite{lee2024echowrist, li2024gazetrak}. To do this, we designed a wristband---placed where skin deformations are most noticeable---containing a speaker and microphone facing towards the forearm and away from the hand. The speaker emits ultrasound waves into the area of the skin encompassing the muscles, and the reflections are then captured by the microphone. The shape of the skin at a given point in time affects the transmission time, reflection angle, and reflection frequency. 
By analyzing reflected signals, we can decode the shape of the deformations over time and estimate how much force is exerted.

\begin{figure}[tb]
  \centering
    \centering
    \includegraphics[width=\linewidth]{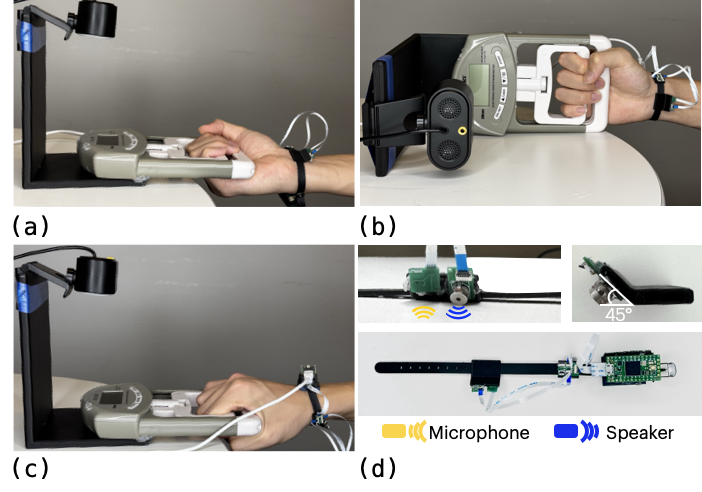}
    \vspace{-2.5em}
    \caption{Study setup, with wrist (a) supinated, (b) neutral, or (c) pronated while exerting grip force on dynamometer. (d) EchoForce hardware: On top, close-up of speaker (blue) and microphone (yellow) modules set at a $45^{\circ}$ downward angle in a 3D-printed bracket. Below, a top-down view of EchoForce.
    }
    \Description{Study setup, with wrist (a) supinated, (b) neutral, or (c) pronated while exerting grip force on dynamometer. (d) EchoForce hardware: On top, close-up of speaker (blue) and microphone (yellow) modules set at a $45^{\circ}$ downward angle in a 3D-printed bracket. Below, a top-down view of EchoForce.
    }
    \label{fig:setup_figure}
    \vspace{-1.8em}
\end{figure}

\subsection{Hardware Prototype} \label{hardware-prototype}

We designed and built EchoForce by integrating custom printed circuit boards (PCBs) with 3D-printed brackets on an off-the-shelf silicone wristband as shown in Figure \ref{fig:setup_figure}d. The wristband was flexible and could be used as a one-size-fits-all design across users by tightening the strap. The bracket containing the speaker and microphone could move around the wristband and be positioned at the center of the wrist when putting on the device. The electronic components on the PCBs include an OWR-0504T-16 speaker and SPH0641LU4H-1 microphone to transmit acoustic signals and capture their reflections, and a Teensy 4.0 Development Board to record and transmit data to a computer. 


As shown in Figure \ref{fig:setup_figure}d, the microphone and speaker were placed on a sensing board mounted on a 3D-printed bracket tilted $45^{\circ}$ away from the skin and facing away from the hand. The angle was chosen based on a pilot study comparing the $45^{\circ}$ and $90^{\circ}$ configurations, where the ML model trained on the $45^{\circ}$ data outperformed the model trained on $90^{\circ}$ by $4.19\%$. The $0^{\circ}$ (flat) configuration was also tested, but deemed unsuitable for evaluation due to frequently shifting in contact with skin during movement. Meanwhile, the primary board with the microcontroller was placed on the opposite side of the wrist to avoid impeding movement. The sensing board was connected to the primary board using a flexible PCB cable.
\subsection{Ground Truth and Data Augmentation}


Grip force was measured using a CAMRY-EH10117 dynamometer \cite{dynamometer} (Fig. \ref{fig:setup_figure}a–c). Past work has validated the reliability of this dynamometer \cite{sanchez-aranda_reliability_2024, wu_grip_2017}. Force was recorded in pounds by the dynamometer, but converted to kilograms in analysis for comparability with past work. To capture a continuous force trace while using the calibration by the dynamometer's microcontroller, we recorded the dynamometer’s digital display at 30 Hz by attaching a camera directly facing the display. However, the device screen updates at approximately 15 Hz, and can freeze the peak value in ``hold'' mode when no force is applied.

In post‐processing, any reading captured while the ``hold'' indicator was active was discarded. Due to missing samples and low refresh rate, we performed data augmentation to obtain a time series with a uniform frequency. Forces during isometric force application typically increase and decay linearly until they plateau at the maximum force \cite{gareis_isometric_1992}. Moreover, most of the missing samples were when the force applied was zero, so we found linear interpolation to be suitable to augment the data. Using this approach, the signal was upsampled to a uniform 100 Hz to match the sampling rate of the internal sensor of the dynamometer. While this made the signal more consistent, it may have compromised the accuracy of ground truth data, which we further explore in Section \ref{limitations}.

\subsection{Signal Processing and Deep Learning} \label{signal_processing}

EchoForce uses Frequency Modulated Continuous Wave (FMCW) signals consistent with past work on acoustic sensing \cite{wangCFMCW2018}. The speaker emits inaudible ultrasonic FMCW waves (ranging from 20 kHz to 29 kHz), and a microphone captures their reflections from skin deformations with a sampling rate of 96kHz. Each frequency sweep consisting of 600 samples constitutes an FMCW frame.

To extract spatial position (reflection delay) and movement information (frequency shift), the collected signal is segmented into FMCW frames. Each frame is cross-correlated with the original transmitted signal, generating a 2D echo profile. In this profile, each column represents the cross-correlation results for a specific frame, encoding both time-domain (horizontal) and spatial-domain (vertical) information. To focus on dynamic movements by offsetting static reflections, a differential echo profile is created by subtracting the previous frame from the current one. These differential echo profiles serve as the input for the deep learning pipeline.


For processing, the 96000 samples per second are divided into 160 frames with a frame length of 600 pixels. 
Meanwhile, the spatial resolution per pixel for our system is 1.79mm\cite{wangCFMCW2018}, so we set the window height to 78 (13.93cm), which captures the tendinous section of the \textit{palmaris longus} muscle. We extract a 320 (2 seconds by 160 frames) by 78 (pixels) moving window for deep learning.

Using Unix timestamps recorded in data collection, echo profiles are synchronized with the ground truth. The ground truth labels from the past two seconds and the extracted moving window from the echo profile compose a single pair of label and input, and the process is repeated for the entire dataset.






We chose FastViT\cite{vasu2023fastvitfasthybridvision} as our encoder model, as it provides the self-attention and long-range dependency processing capabilities of vision transformers \cite{VisionTransformer}, with quick training. For decoding, we used an average pooling layer, a dropout layer (0.8 dropout probability), and a fully connected layer in series. We used root mean squared error (RMSE) as the evaluation metric due to its use in past work on force estimation \cite{keir_development_2005}, whereas we used mean squared error (MSE) as the loss function. We used SGD as the optimizer with 0.001 weight decay, 0.001 initial learning rate, and 60 batch size.

%% file: 04_study.tex
\section{Study Design}  \label{study_design}
We conducted a study to evaluate continuous grip force estimation using our system, focusing on its ability to generalize between sessions, hand orientations, and users. Our study was approved by the Institutional Review Board (IRB) at Cornell University.


\subsection{Measures and Task}
Maximum voluntary contractions (MVC) indicate the highest level of force generated by a muscle group, and percentiles of these are a standard form of evaluation in EMG studies \cite{keir_development_2005,mogk_prediction_2006}. We used MVCs to define the levels of force to be applied by each participant. To report relative error rate in a consistent manner with past work \cite{keir_development_2005,liu_fingertip_2014}, we report results as a percentage of each participant's MVC, following the equation: \textbf{$RMSE / MVC  \times 100= Error Rate (\%MVC)$}, hereafter reported as \textbf{error rate}. Error rate is particularly important to account for variance in participant strength and ensure comparability with different participant populations between studies.

Each participant’s MVC was first measured and used to compute five target levels corresponding to 5\%, 25\%, 50\%, 75\%, and 100\% of that maximum. During each trial, one target level was displayed on the laptop (both in percentage and absolute force) in random order, and participants were instructed to increase their force to match the displayed percentage as quickly and accurately as possible, and then release. Participants also heard the target level via text-to-speech audio to assist in pacing. Participants could view the amount of force currently being applied on the dynamometer as feedback.

Each participant completed five sessions of data collection in each of the three wrist orientations: supinated, neutral, and pronated (shown in Figure \ref{fig:setup_figure}(a), (b), and (c), respectively). In each session, every target level was displayed for three seconds in a random order, followed by a one-second rest period, composing one trial. However, participants completed the grip-and-release action in two seconds on average. This procedure yielded six repetitions of each target level per session. Altogether, this resulted in 450 trials per participant, 150 trials per orientation, and 30 trials per session.

\subsection{Study Procedure and Participants}
The procedure of the user study was as follows:
\textbf{(1) Introduction:} Participants signed the informed consent form, completed a brief demographic survey, and received an overview of the experiment. \textbf{(2) Practice Session:} After reviewing study structure, each participant performed a session in the pronated orientation; no data was recorded. \textbf{(3) MVC Measurement:} We measured the participant’s MVC. \textbf{(4) Data Collection:} A two-minute recording was conducted in the current hand orientation, comprising 1 of the 15 total sessions recorded. \textbf{(5) Remount:} The device was fully removed and remounted in less than ten seconds. \textbf{(6) Repeat:} Steps 4 and 5 were carried out five times for each of the three orientations (fifteen sessions in total). The total duration of the study was 60 minutes (including introduction and surveys). Upon
completion of the study, each participant received a $\$20$ gift card as compensation.

We recruited 12 participants. However, we dropped one participant’s data due to data loss during the user study. 
Consequently, we analyzed data from 11 participants (8 male, 2 female, and 1 preferred not to respond) aged 19-33 (MD=23, SD=4.09), and all but one participant self-reported to be right-handed. However, all participants performed the experiment with their right hand to keep the protocol and data consistent in the experiment, especially for user-independent evaluation. All participants were undergraduate and graduate students in the same institution. Average MVC among participants was 26.14kg, significantly lower than typical grip strength for the 18-29 age group (47kg for men, 28.1kg for women) \cite{wangHandGripStrengthNormative2018}. Section \ref{limitations} discusses potential causes.

%% file: 05_results.tex
\section{Results}


We trained four types of models to evaluate EchoForce. \textbf{User-dependent} models for testing the ability to generalize to a new session; \textbf{user-independent} models for testing the ability to generalize to new users; \textbf{fine-tuned} models for testing the usefulness of data from multiple users in user-dependent testing; and \textbf{cross-orientation} models for testing the ability to generalize to new orientations. For all results, the RMSE in kg and mean error rate in \%MVC for each participant are summarized in Table \ref{tab:error_comparison}. We calculated \textbf{error rate in terms of \%MVC as RMSE divided by the maximum voluntary contraction (MVC) for each participant}.

\subsection{Force Estimation}

\subsubsection{User-Dependent Results}

In our user-dependent testing, with 15 sessions ($5 \times 3$ wrist orientations) per participant, we left the last session out for each wrist orientation to be part of the testing dataset and trained the user-dependent regression model for each participant on the remaining 12 sessions. We trained the model for 40 epochs for each participant based on pilot testing.

In our user-dependent evaluation, we found that the model achieved 10.10\% (2.56kg) mean error rate. The error ranged from 1.75kg to 3.5kg between participants. For error rate, results ranged from 6.01\% to 14.34\%, consistent with results in EMG-based studies that use a similar protocol \cite{keir_development_2005}. As the device was remounted between each session, these findings demonstrated that EchoForce can generalize to sessions for which no data were included in training.

\begin{table*}[ht!]
  \centering
  \caption{Comparison of Error Metrics (kg) and Error Rates (\%MVC) Across Different Methods}
  \Description{Comparison of Error Metrics (kg) and Error Rates (\%MVC) Across Different Methods}
  \vspace{-0.8em}
  \label{tab:error_comparison}
  \renewcommand{\arraystretch}{0.92}
  \setlength{\tabcolsep}{1pt}
  \small
  \begin{tabular*}{\textwidth}{@{\extracolsep{\fill}} l c *{12}{c}}
    \toprule
    \multicolumn{2}{c}{User ID} & {P1} & {P2} & {P3} & {P4} & {P5} & {P6} & {P7} & {P8} & {P9} & {P10} & {P11} & {\textbf{Mean}} \\
    \midrule
    \addlinespace
    \multicolumn{2}{c}{Max. Voluntary Contraction (kg)}  & 27.00 & 22.05 & 31.50 & 19.80 & 18.00 & 30.15 & 27.00 & 27.90 & 29.25 & 31.95 & 22.95 & \textbf{26.14}  \\
    \midrule
    \addlinespace
    \multirow{2}{*}{User-Dependent}
    & {RMSE (kg)} & 1.75  & 2.44  & 2.14  & 1.94  & 2.46  & 3.02  & 3.50  & 2.44  & 3.29  & 1.92  & 3.29  & \textbf{2.56}    \\
    & Error Rate (\%MVC) & 6.48  & 11.07 & 6.79  & 9.80  & 13.67 & 10.02 & 12.96 & 8.75  & 11.25 & 6.01  & 14.34 & \textbf{10.10}   \\
    \addlinespace
    \multirow{2}{*}{User-Independent}
    & {RMSE (kg)} & 2.44  & 3.54  & 2.89  & 3.17  & 2.43  & 4.04  & 3.19  & 2.62  & 3.39  & 2.60  & 3.93  & \textbf{3.11}    \\
    & Error Rate (\%MVC) & 9.04  & 16.05 & 9.17  & 16.01 & 13.50 & 13.40 & 11.81 & 9.39  & 11.59 & 8.14  & 17.12 & \textbf{12.29}   \\
    \addlinespace
    \multirow{2}{*}{Fine-Tuned}
    & {RMSE (kg)} & 1.66  & 2.25  & 2.18  & 2.03  & 1.98  & 2.56  & 2.81  & 2.13  & 3.33  & 1.72  & 2.78  & \textbf{2.31}    \\
    & Error Rate (\%MVC) & 6.15  & 10.20 & 6.92  & 10.25 & 11.00 & 8.49  & 10.41 & 7.63  & 11.38 & 5.38  & 12.11 & \textbf{9.08}    \\
    \addlinespace
    \multirow{2}{*}{Cross-Orientation}
    & {RMSE (kg)} & 3.76 & 3.48 & 2.93 & 2.71 & 2.43 & 4.52 & 4.08 & 2.56 & 3.9 & 2.64 & 3.53 & \textbf{3.32} \\
    & Error Rate (\%MVC) & 13.94 & 15.78 & 9.31 & 13.67 & 13.48 & 14.98 & 15.1 & 9.17 & 13.33 & 8.25 & 15.37 & \textbf{12.94} \\
    \addlinespace
    \bottomrule
  \end{tabular*}
  \vspace{-1.6em}
\end{table*}


\begin{figure*}[htb]
  \centering
    \centering
    \includegraphics[width=0.95\linewidth]{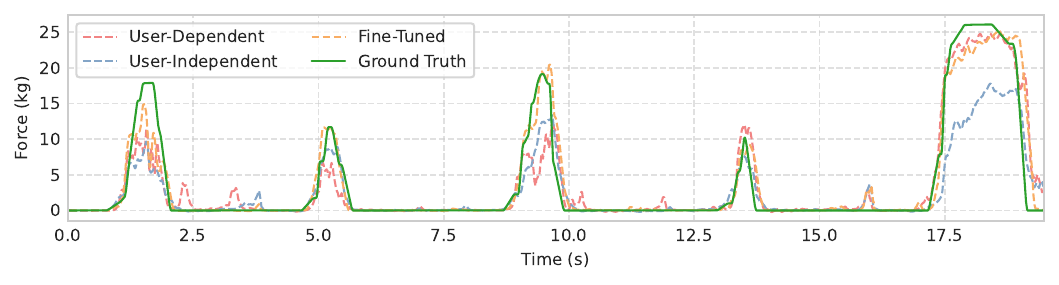}
    \vspace{-1.6em}
    \caption{Predictions for Participant 1 with a five-point moving-average filter applied for clarity.}
    \Description{Predictions for Participant 1 with a five-point moving-average filter applied for clarity.}
    \label{fig:time_series}
    \vspace{-1em}
\end{figure*}


\subsubsection{User-Independent Results}

To demonstrate the generalizability to new users, we conducted user-independent, leave-one-user-out testing. We evaluated each model by leaving out one user for testing and training on the remaining participants. The user-independent models were each trained for 10 epochs. 

For user-independent, leave-one-user-out evaluation, we achieved 12.29\% (3.11kg) mean error rate. The user-independent model only performed slightly worse than the user-dependent, ranging from 2.43 kg to 4.04 kg, while the mean error is 0.55 kg larger. Considering the high variance in muscle activity, as well as hand and arm shape between people, this was an impressively accurate result, overcoming a major limitation of past work on grip force estimation. These results demonstrate that EchoForce can be effective without needing to collect training data or perform calibration for a new user, drastically reducing setup time and making EchoForce suitable for practical real-world use.



\subsubsection{Fine-Tuned Results}

To evaluate whether user-dependent results can be further enhanced through data from additional users, we used fine-tuning. Starting with the user-independent models, we fine-tuned them with the training set of the participants' own data. The training hyperparameters and the dataset are the same as the user-dependent models, except that the fine-tuned model is initialized with the pre-trained user-independent model.

For fine-tuned, user-dependent evaluation, we achieved 9.08\% (2.31kg) mean error rate. Compared to randomly initialized user-dependent training, fine-tuning decreases mean RMSE by 1.02\% (0.25 kg). These results demonstrate that a larger dataset would significantly enhance the user-dependent accuracy of EchoForce. Therefore, for applications where highly accurate continuous force measurement is needed, EchoForce could provide a pre-trained model based on a large dataset, and further improve the accuracy by fine-tuning with the user's data if needed.


Figure \ref{fig:time_series} shows the force traces for the ground truth and regression predictions from user-dependent, user-independent, and fine-tuned models over time for Participant 1. As shown, our system can identify the onset time of the force, make predictions about the trend, and continuously estimate grip force. Furthermore, the fine-tuned model clearly outperforms user-dependent and user-independent models, reducing false positives and predicting the magnitude more accurately.

\subsubsection{Within-User Cross-Orientation Results}

Prior user-dependent models were trained on data from all three wrist orientations. To validate our system's ability to generalize to new hand poses and orientations, we conducted a cross-orientation evaluation. For each participant, we performed leave-one-out testing with one orientation used for testing and the other two orientations for training.

For our user-dependent, cross-orientation model, we achieved 12.94\% (3.32 kg) mean error rate. While this was worse than other user-dependent results, it included significantly less of each user's data (66.7\% vs. 80\%) for training. The mean RMSEs of orientations (``pronated'', ``neutral'', and ``supinated'') for participants are 3.5 kg, 3.44 kg, and 3.02 kg, and the mean error rate is 13.63\%, 13.43\%, and 11.77\%. While there was no major difference, the orientation with highest accuracy is supinated, where the palm is facing upwards. The cross-configuration results indicate the system can predict grip force with orientations that are not in the training dataset.

\subsection{Usability}

At the end of the study, participants rated the device’s comfort on 5-Likert scale ranging from 1 (“very uncomfortable”) to 5 (“very comfortable”). The mean score was 4.63 (SD = 0.50), indicating that EchoForce was comfortable throughout the study. All participants indicated not hearing any audible sound emitted from the device.

%% file: 06_discussion.tex
\section{Discussion}


EchoForce sought to use a novel acoustic sensing approach to develop a cost-effective system that is easy to put on or integrate into other devices, while overcoming limitations of past methods, such as the need to recalibrate models for each session or user.

\subsection{Findings and Comparison}

EchoForce matched EMG-based approaches in user-dependent conditions, achieving marginally better results at 9.08\% error rate compared to Keir and Mogk, who found 11.4\% error rate when estimating grip force using EMG with a similar maximal grip force protocol \cite{keir_development_2005}. Meanwhile, it drastically outperformed in user-independent testing, where even simplified ``static'' calibration by Hoozemans et al. resulted in 6.1kg error \cite{hoozemans_prediction_2005}. Similarly, light reflectivity approaches such as HIPPO drastically increased in error from 4.2kg to 9.72kg when no fine tuning is performed for new users \cite{yin_hippo_2023}. Hoozemans et al. and HIPPO did not report relative error rate, so we used RMSE for comparison. While EchoForce had higher error rate in cross-configuration testing, it still outperformed past work on sensing grip force in multiple degrees of freedom, which had a 16\% error rate even when all orientations are included in training \cite{nielsen_simultaneous_2011}. Overall, our results demonstrate the adaptability of EchoForce to new data with invariance to session, user, and hand orientation.

\subsection{Practical Considerations}

EchoForce shows reliable grip force estimation with high comfort, thanks to its compact sensing components (12 $\times$ 7.3 mm speaker and 8.6 $\times$ 8.6 mm microphone) using active acoustic sensing. This system design enables low-power operation \cite{lee2024echowrist, li2024eyeecho} without bulky hardware, facilitating the future integration of EchoForce into real-world wearable devices such as wristbands or smartwatches.



Privacy is crucial for the real-world adoption of acoustic sensing technologies. EchoForce addresses this concern by operating in the inaudible ultrasound range, ensuring that speech and other audible sounds can be filtered out during signal processing. To further enhance privacy, this processing would ideally be performed on-device when deployed outside the laboratory setting.



EchoForce successfully tracks grip force, but its performance in dynamic scenarios, e.g., walking or wrist motion, remains unexplored. While these movements may introduce artifacts, we are optimistic about mitigation. Our cross-orientation robustness suggests that ML can suppress noise from different poses. Future work can further learn the motion patterns and apply motion-adaptive calibration or filtering to remove artifacts. Following prior acoustic sensing work \cite{li2024eyeecho, zhang2023echospeech}, future research should evaluate performance in dynamic conditions (e.g., daily activities).

In this research, we focus on optimizing grip force tracking on a wristband. As detailed in Section~\ref{hardware-prototype}, since sensor placement (e.g., location and angle) impacts the performance, we developed a customized device to achieve optimal results, rather than relying on off-the-shelf solutions. For instance, the speaker and microphone configurations in commercial smartwatches are often positioned on opposing sides and pointing in different directions, capturing more extraneous noise. Future work will explore methods to mitigate noise induced by suboptimal sensor placement, thereby enabling robust grip force tracking on off-the-shelf devices.

\subsection{Applications}
EchoForce demonstrates initial feasibility of acoustic grip force measurement, paving the way for future applications in health and HCI. For instance, it can monitor grip force continuously with minimal setup, making it highly suitable for health monitoring and rehabilitation, as traditional methods like EMG are often cost-prohibitive, time-consuming, and require clinical expertise. In particular, we plan to further evaluate its application for post-stroke paresis (i.e. muscle weakness) where patients often struggle to \textit{sustain} grip force during daily activities, making continuous measurements important for monitoring recovery \cite{kamimuraEvaluationSustainedGrip2002}. For HCI, EchoForce offers a practical method to enable force-based input for wristbands or smartwatches, providing an intuitive means of continuous control in a wide range of environments \cite{mollyn_egotouch_2024, buddhika_fsense_2019}. Moreover, the sensing method of EchoForce could be adapted to estimate muscle activity, either directly through skin deformations or indirectly via grip force, which can reliably reflect muscle activity in the wrist \cite{mogk_prediction_2006}.

\subsection{Limitations}
\label{limitations}
While the performance and applicability of EchoForce are promising, several key limitations should be discussed.
First, the study was conducted in a controlled lab setting with limited extraneous movements. While our models were able to successfully generalize between sessions, users, and hand orientation, the hand is highly flexible and can perform many movements beyond what was included in this study. Moreover, the current system is not suitable for capturing activity from the extensor muscles at the back of the palm, but these are not tensed during grip or grasping actions. As all users in our study were able-bodied and healthy, we also do not know how well the system would perform for users with myopathy, which is valuable for rehabilitation or health monitoring.

Our ground truth measurements were based on computer vision-based tracking of a dynamometer screen. While this made it convenient to rely on the internal calibration of the dynamometer, it resulted in several dropped frames, which needed to be estimated by linear interpolation. This interpolation may have compromised the fidelity and accuracy of our ground truth. In future work, this problem can be resolved by using more advanced dynamometers that can directly stream data which would be more suitable for precise, continuous monitoring of grip force over time \cite{biometricHandGripDyno}). However, holding a dynamometer can limit the grasp types used for applying force. We plan to explore hand interactions with a variety of different objects, as done in past studies \cite{fan_what_2018}.

Participants in our study had lower grip strength than expected for their demographics. We believe that this may have been caused by our use of a simplified version of the  ASHT grip strength protocol based on prior work \cite{keir_development_2005,hoozemans_prediction_2005}, and due to our particular graduate student population, which may not be representative of the general population. For future work, recruiting a more diverse group of participants is needed. Moreover, there was high variance in accuracy between participants, especially when compared to their MVCs. Our system typically performed better for users with a higher MVC, likely because they have more pronounced skin deformations when applying higher forces. These observations suggest that muscle mass or body weight may impact system performance. Future work should examine how factors such as weight, height, body fat percentage, and wrist circumference affect results. In the present study, these effects were partially mitigated in our fine-tuned model, where users with a high error rate were reduced to be closer to the mean. For example, P11's error rate was reduced from 14.34\% to 12.11\%. Therefore, this limitation may be overcome with data from more users with greater body diversity.

%% file: 07_conclusion.tex
\section{Conclusion}
EchoForce is a wristband using a novel application of acoustic sensing to continuously measure grip force via subtle skin deformations using only a speaker and microphone. We evaluated EchoForce with 11 participants, achieving a 9.08\% user-dependent error rate and a 12.3\% user-independent error rate. EchoForce maintained accuracy independent of remounting sessions, users, and hand orientations, overcoming major limitations of prior force sensing approaches. EchoForce is inexpensive, does not need calibration, and can be worn and taken off quickly, making it highly practical for wearable grip force measurement in health monitoring and HCI.

%% file: 08_acknowledgements.tex
\begin{acks}
This project was partially supported by the National Science Foundation Grant Award No. 2239569.
\end{acks}